\documentclass [12pt,preprint]{aastex}
\begin{document}

\title{Modelling Observational Constraints for Dark Matter Halos}

\author{F.D.A. Hartwick}

\affil{Department of Physics and Astronomy, \linebreak University of 
Victoria,
Victoria, BC, Canada, V8W 3P6}
\begin {abstract}

Observations show that the underlying rotation curves at intermediate radii in
spiral and low-surface brightness galaxies are nearly universal. Further, in 
these same galaxies, the product of the central density and the core radius 
($\rho_{0}r_{0}$) is constant. An empirically motivated model for dark matter 
halos which incorporates these observational constraints is presented and 
shown to be in accord with the observations. A model fit to the observations 
of the galaxy cluster Abell 611 shows that $\rho_{0}r_{0}$ for the dark matter
halo in this more massive structure is larger by a factor of $\sim 20$ over 
that assumed for the galaxies. The model maintains the successful NFW form in 
the outer regions although the well defined differences in the inner regions 
suggest that modifications to the standard CDM picture are required.

\end{abstract}


\section {Introduction}

Recent observations suggest that many spiral galaxies possess constant density 
cores whose central densities are related to their core radii. In addition 
these same galaxies possess nearly universal rotation curves at intermediate 
radii after removal of the baryonic component. These observations are 
used to construct 
a simple empirical model for the density distribution of dark matter halos. 
This is done without prejudice for any specific model of structure formation 
although the model is designed to retain the sucessful NFW form (Navarro et 
al. 1997) in the outer regions.

For many years we have relied on sophisticated cosmological simulations to 
provide guidance on the distribution of dark matter (hence the NFW profile). 
Only relatively 
recently has {\it{observational}} data on the distribution of dark matter over 
a large range in radius become available (e.g. the data on the galaxy cluster 
Abell 611 from Newman et al. 2009). If the empirical model derived from 
the galaxy sample above is to describe all dark matter halos and hence to 
provide constraints on the properties of 
the dark matter particle itself then it is important to test it on as many 
scales and morphologies as possible. For this reason we include an 
application of our model to the observations of the galaxy cluster Abell 611.

\section{The Model}

The model is motivated by two independent observational results. The first is 
from Donato et al. (2009) who showed that the product of the central dark 
matter density and its core radius (the radius within which the density 
remains approximately constant) is a constant with a value of $\sim120 M_
{\odot}pc^{-2}$. Salucci et al. (2011) emphasize that this result 
extends to the dwarf spheroidal galaxies as well. This result builds  
upon earlier work by Kormendy \& Freeman (2004). The second observational 
result is from the work of McGaugh et al. (2007). Here an apparently 
universal relation between rotational velocity and radius at intermediate 
radii in 60 spiral galaxies is found after removal of the contributions from 
the stars and gas. Walker et al. (2010) then extend the 
same relation to dwarf spheroidal and low-surface-brightness (LSB) galaxies.

In order to account for the above observational results the following density 
profile is proposed
\begin{equation}
\rho(r)=\rho_{0}/(1+r/r_{c})(1+r/r_{s})^{2}
\end{equation}
where $\rho_{0}$ is the constant central density, $r_{c}$ is a scale factor, 
and $r_{s}$ is a second scale factor and is defined as $f\times r_{c}$ where 
$f > 1$. The core radius 
($r_{0}$) is defined as that radius where the density has fallen to 
$\rho_{0}/4$. (This is the definition of core radius from the Burkert (1995) 
profile). The observations show that $\rho_{0}r_{0}$ is a constant so that for
a given assumed value of $\rho_{0}$ and $f$, $r_{c}$ found from a solution 
of the following cubic equation will satisfy this constraint.
\begin{equation}
r_{c}^{3}-((2/f+1)r_{0}/3)~r_{c}^{2}-((2/f+1/f^{2})r_{0}^{2}/3)~r_{c}-r_
{0}^{3}/3f^{2}=0
\end{equation}
Note that with $\rho_{0}r_{0}$ given by the observations the problem contains 
only the two free parameters $\rho_{0}$ and $f$.

The mass enclosed within radius $r$ is given by
\begin{equation}
M(r,a,b)=4\pi\rho_{0}(1/(b^{2}(a-b)(br+1))+((a-2b)\ln (br+1)/(b(a-b))^{2}+\ln 
(ar+1)/(a(a-b)^{2})-1/(b^{2}(a-b)))
\end{equation}
where $a=1/r_{c}$ and $b=1/r_{s}$.

The outer boundary of the model occurs when $M/(4\pi r^{3}/3)=200\rho_{crit}$ 
($h=0.7$ assumed).
Given that this boundary condition is the same as that assumed for standard 
CDM, the predicted baryonic Tully-Fisher relation here is also the same (i.e. 
$M_{b}\propto V^{3}$), (e.g. Steinmetz \& Navarro 1999).

In order to tie the above density distribution to a solution of the Jean's 
equation we use the pseudo-phase-space density $Q=\rho/\sigma_{r}^{3}$ as the  
second variable. This variable has interesting properties which could be 
relevant to the structure formation problem (Hogan \& Dalcanton, 2000; 
Dalcanton \& Hogan , 2001). $Q$ 
was shown by Taylor and Navarro (2001) to vary with 
radius as $r^{-1.875}$ in the NFW model. In order to maintain 
this behaviour in the outer region of the above model and to keep the problem  
simple we assume the following form for the radial dependence of $Q$
\begin{equation}
Q=Q_{0}/(1+(r/r_{Q})^{1.875})
\end{equation}
With the spherical Jean's equation written in terms of $\rho$ and $Q$ 
(Hartwick, 2008) 
and the above differentiable analytical expressions for both variables the 
solution is the 
run of the anisotropy parameter $\beta$ (Binney \& Tremaine, 1987) i.e.
\begin{equation}
1.2\beta = -0.6\frac{GM_{r}}{r}\left(\frac{Q}{\rho}\right)^{2/3}-\frac{dlog \ 
\rho}{dlog \ r} +0.4\frac{dlog \ Q}{dlog \ r}
\end{equation}
with $Q_{0}$ and $r_{Q}$ as free paramters (for fixed $\rho_{0}$ and $f$). By 
choosing $r_{Q}=r_{0}$ and 
varying $Q_{0}$ until $\beta \sim 0$ at $r_{200}$ solutions were found where 
$\beta$ rose smoothly from zero at the center to a maximum of 
$\sim 0.35$ followed by a not so smooth descent to zero at the outer boundary 
(e.g. Navarro et al. 2010, Fig. 10). Adding an extra parameter to (4) (e.g. 
equ'n (1) of Hartwick, 2008) results in some smoothing and changes $Q_{0}$ by
$\sim \pm 20\%$ even if $Q$ is allowed to remain at or higher than its maximum 
value at the outer boundary. Ultimately the observations must show the true 
behaviour of the run of $\beta$. $Q_{0}$ was determined by iteration using the
secant method. 
 
Under the assumption that $\rho_{0}r_{0}=120$ $ M_{\odot}pc^{-2}$ as $f$ 
approaches unity a minimum mass 
is found for each value of $\rho_{0}$ assumed. In the 
case of $\rho_{0}=5.0$ $M_{\odot}pc^{-3}$ , for example, this minimum is 
$\log M_{200} =7.15$ $ M_{\odot}$ with $Q_{0}=9.8\times 10^{-3}$ $M_{\odot}
pc^{-3}(km \ sec^{-1})^{-3}$. As the mass is increased at fixed central 
density (by increasing $f$) $Q_{0}$ slowly decreases. At $\rho_{0}=0.01$ 
$M_{\odot}pc^{-3}$ this minimum mass increases to $\log M_{200} =12.08$ $ M_
{\odot}$ with $Q_{0}=2.4\times 
10^{-9}$ $M_{\odot}pc^{-3}(km \ sec^{-1})^{-3}$. More solutions are shown in 
the next section where they are compared to the observations.

\section{Comparison with the Observations}

In this section we show a comparison between the model and a sample of 
observations. 
Unless otherwise stated the models were computed assuming that $\rho_{0}r_{0}=
120~M_{\odot}pc^{-2}$.

\subsection{The Fornax Dwarf Spheroidal}

Figure 1 shows the results of several analyses of the stellar kinematics 
of Fornax. Plotted in the $log M - log R$ plane are results from Strigari 
et al. (2007); Strigari et al. (2008); Walker et al. (2010) \& Amorisco \& 
Evans (2011) plotted as filled circles. The two open circles are from the work
of Walker \& Penarrubia (2011) whose analysis makes use of two independent 
stellar populations within the galaxy. Parameters of the model shown (solid 
line) are $\rho_{0}=0.24~M_{\odot}pc^{-3}, f=2.0,~Q_{0}=5.3\times10^{-6}~M_
{\odot}pc^{-3}(km \ sec^{-1})^{-3}~ \&~ \log M_{200}=9.66~M_{\odot}$.

\subsection{Spiral and LSB Galaxies}

Figures 2 \& 3 show the models superimposed on the observations of McGaugh et 
al. (2007) of spiral galaxies (open circles) and a sample of LSB galaxies 
(crosses) Kuzio de Naray et al. (2010). The spiral galaxy sample has had the 
mass in stars and gas removed. Figure 2 shows 4 models with density 
$\rho_{0}=0.3~M_{\odot}pc^{-3}$ and values of $V_{200}~(kmsec^{-1})$ indicated
. This is to be compared to the upper 
Figure 3 of McGaugh et al. (2007) where NFW models are superimposed on the 
same data. Note how the relation between concentration index (c) and mass 
cause each NFW solution to be separated from the other whereas the solutions 
here are approximately asymptotic about the solid black line. This 
line shown extended in Fig. 3 is a least squares fit to the spiral data alone 
and has a logarithmic slope of 0.5 (McGaugh et al. 2007).  
Figure 3 shows the same data superimposed 
by 3 models all with $M_{200}=3\times10^{12}~M_{\odot}$ but different central 
densities (red, 5.0; green, 0.1 \& blue, $0.01~M_{\odot}pc^{-3}$) which 
illustrates the point that models with the same mass can have different dark 
matter distributions.

\subsection{Clusters of Galaxies}

Up to this point, all model results have been obtained by assuming 
$\rho_{0}r_{0}=
120~M_{\odot}pc^{-2}$. This assumption appears to break down when attempting 
to fit the model to the observations of the galaxy cluster Abell 611 
(Newman et al. 
2009). These authors used a combination of weak and strong lensing along with 
spectroscopy of the central galaxy to obtain the dark matter distribution over 
a very large range in radius. In order to obtain a reasonable fit to the above 
observations, the model required that $\rho_{0}r_{0}=2350\pm 200~M_{\odot}pc^
{-2}$ (i.e. roughly $20$ times the value assumed for the galaxies). 

Recently Newman et al. (2012) have presented the results of a similar 
analysis for 7 massive clusters including Abell 611. The authors fit to 
each cluster a 
density profile similar to equ'n (1) but with a different parameterization. 
Their paper does not contain enough information to compute individual values 
of $\rho_{0}r_{0}$. However from their Fig. 3, we estimate that the central 
densities are $\sim 0.1-0.2~M_{\odot}pc^{-3}$ and with mean core sizes (their 
$r_{c}$ generally will be less 
than our $r_{0}$) of $\sim 13 kpc$ $\rho_{0}r_{0}$ appears to be at 
least 10 times higher than the galaxy value. Observations of 
intermediate mass structures are needed to determine whether the relation
between $\rho_{0}r_{0}$ and dark halo mass is continuous.

Our model fit to Abell 611 is shown in Figure 4 as the red line. 
The model parameters are $\rho_{0}=0.1~M_{\odot}pc^{-3},
f=9.0,~Q_{0}=6.7\times10^{-11}~M_{\odot}pc^{-3}(km \ sec^{-1})^{-3}$ and  
a mass $log M_{200}=14.57~M_{\odot}$.

\section{Summary}

The observations show that many spiral, and LSB galaxies exhibit a nearly 
universal relation between the underlying rotational velocity and radius at 
intermediate radii and a relation between the central density and core radius 
such that their product is a constant. A model is constructed based 
on these contraints and results are shown to be consistent with the 
observations and to differ in the inner region from predictions based on the 
NFW profile of CDM. Specifically, in addition to the galaxies possessing a dark
matter 
core, the mass-concentration relation appears to be absent and for central 
densities above a mass dependent threshold density 
dark halos of the same total mass can have different internal structures i.e. 
larger central densities and smaller core radii.

The model was also fit to the observations of the galaxy cluster Abell 611. 
It was found that the central density-core radius relation had to be $\sim 20 
$ times higher than that found for the galaxy sample. A recent study of the 
density profile of 7 massive clusters by Newman et al. (2012) also suggests a 
significantly higher value. Additional observations are required to determine 
if $\rho_{0}r_{0}$ varies continuously between the masses of the dark matter 
halos of the galaxies and those of the galaxy clusters. 
 
The model provides an empirically-motivated cored template which may be 
useful for disentangling the relative contributions of dark matter and  
baryons in other types of galaxies such as the ellipticals.

Should the observational constraints which define the model be 
shown to be compatible with other galaxy types (e.g. the ellipticals) 
then accounting for them will strengthen the need for major modifications to 
the standard CDM picture. 

One way to modify CDM is to invoke the idea that 
the dark matter particles are self interacting (e.g. Spergel \& Steinhardt 
2000). If the mean free path of the particles is of the order of the core size 
$r_{0}$, then the scattering coefficient will be $\sim (\rho_{0}r_{0})^{-1}$. 
In the case of Abell 611 this quantity in cgs units is $2.0\pm 0.2~cm^{2}gm^
{-1}$. However to be consistent with the value for the galaxies, $\sim 40.0~cm^
{2}gm^{-1}$, the scattering coefficient could be energy dependent  
given the difference in mass between the galaxies and the clusters. Such
an energy dependent scattering coefficient has been investigated by 
Loeb \& Weiner (2011). A further consideration is that if the coefficient is 
sufficiently large the halo is subject to core collapse within a Hubble 
time (e.g. Burkert 2000). Clearly, many uncertainties with this scheme remain.

The self interacting dark mattter (SIDM) picture mentioned above is 
but one of several which have been proposed to explain the presence 
of cores in dark matter halos. An up-to-date list is discussed in the paper 
by Newman et al. (2012). The challenge for any theory is to predict the 
observed relation between $\rho_{0}$ and $r_{0}$. The continuing observational 
program of separating the distribution baryons and dark matter on all scales 
and for all morphological structures should eventually lead to the correct 
scenario.

\acknowledgements

The author wishes to thank Andrew Newman and Richard Ellis for kindly 
providing the data from Newman et al. (2009) on the galaxy cluster Abell 611 
in the form 
shown in Fig 4. He also gratefully acknowledges
support for this work from a discovery grant from NSERC (Canada).

\clearpage

\begin{figure}
\plotone{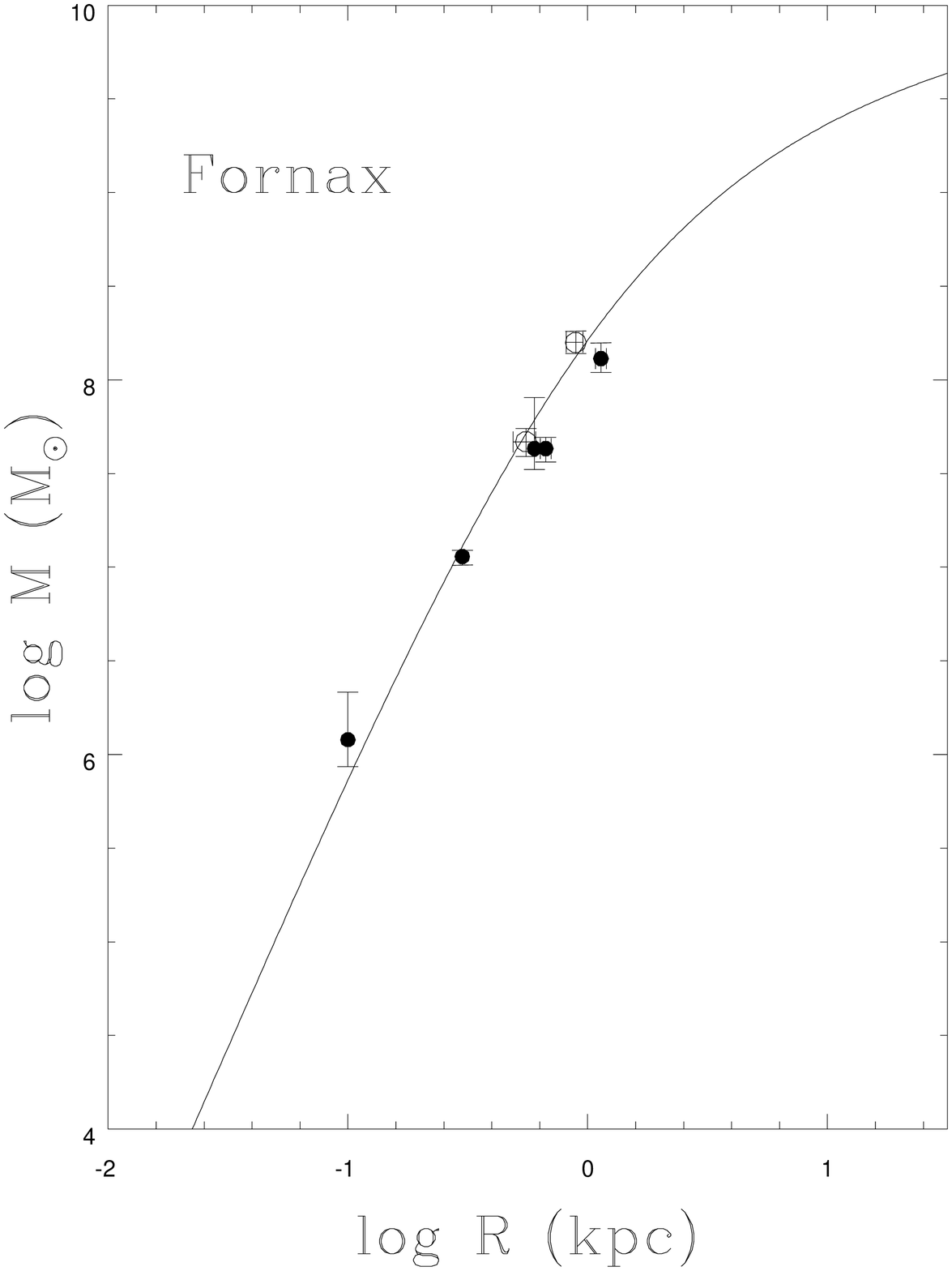}
\caption{A model with parameters $\rho_{0}=0.24,~f=2,~Q_{0}=5.3\times 10^{-6}
~\&~log M_{200}=9.66~M_{\odot}$ fit to the modelled stellar kinematical data 
cited in the text for the Fornax Dwarf Spheroidal Galaxy} 
\end{figure}

\clearpage

\begin{figure}
\plotone{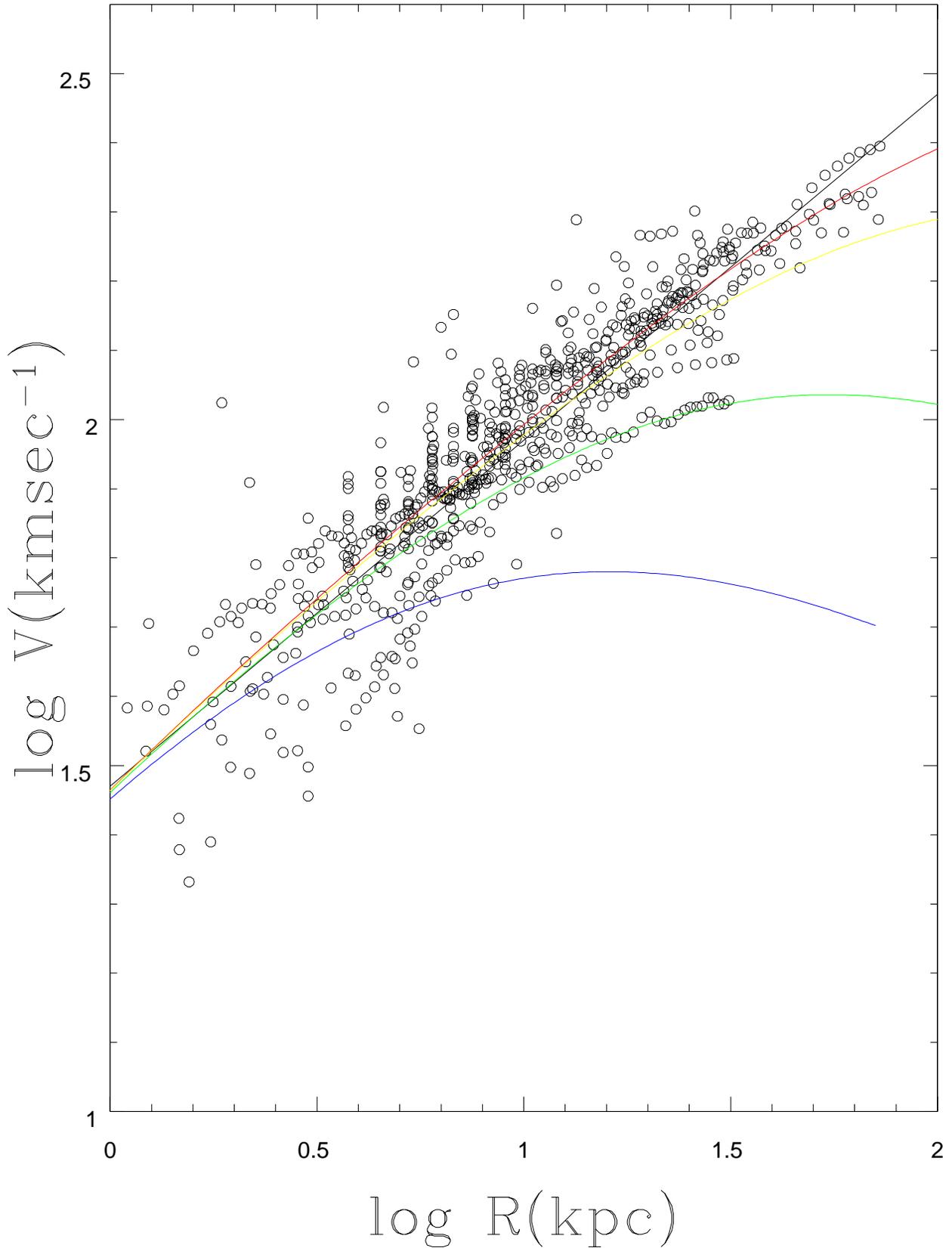}
\caption{Four models superimposed on the data from fig 3 of McGaugh et al. 
(2007). All models have $\rho_{0}=0.3$ but with $V_{200}=300,Q_{0}=7.1\times 
10^{-7}$ (red); $V_{200}=200,Q_{0}=1.1\times 10^{-6}$ (yellow); $V_{200}=100,
Q_{0}=2.2\times 10^{-6}$ (green); \& $V_{200}=50,Q_{0}=4.3\times 10^{-6}$ 
(blue)}
\end{figure}

\clearpage

\begin{figure}
\plotone{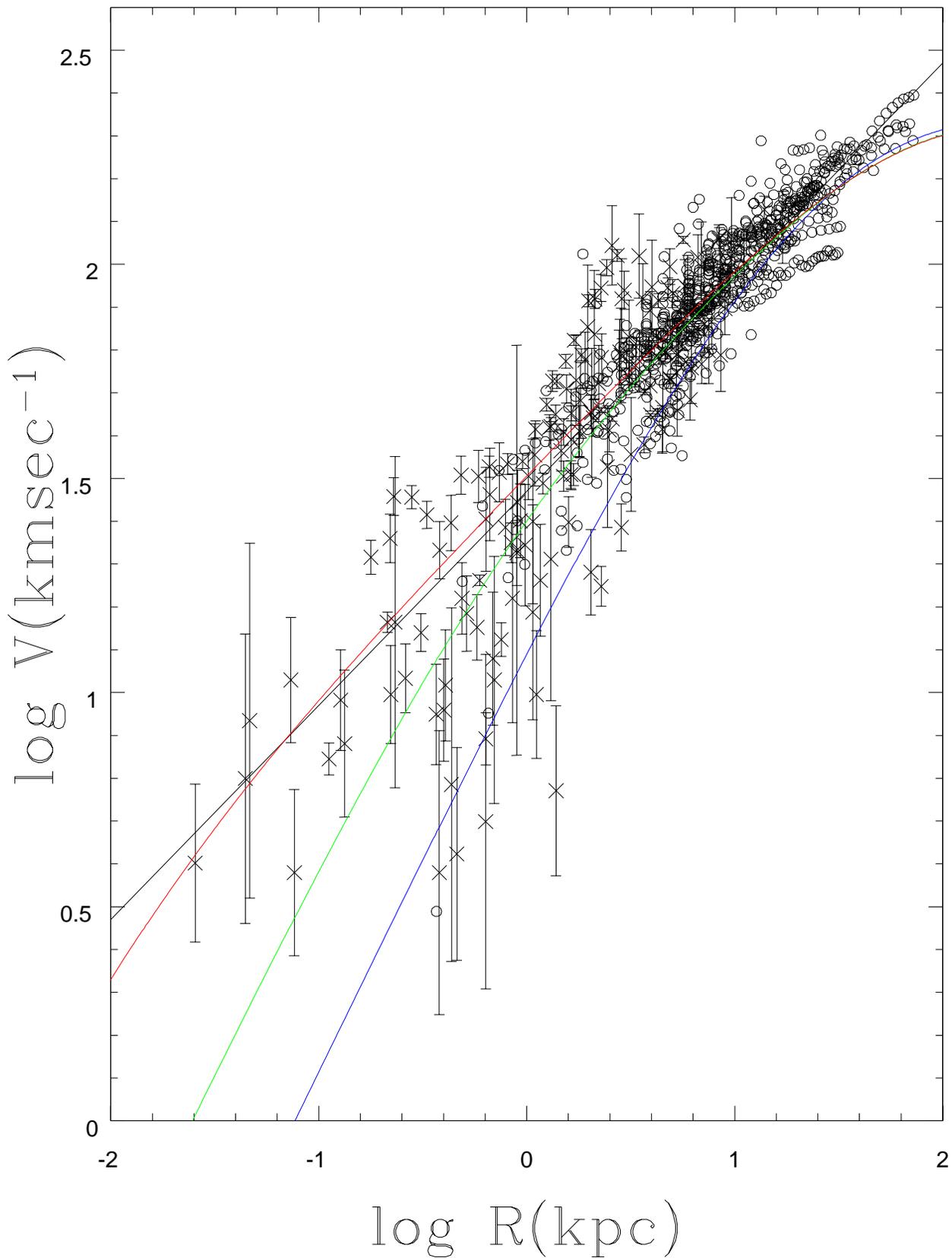}
\caption{Three models superimposed on the data from fig 3 of McGaugh et al. 
(2007) extended to include data for LSB galaxies from Kuzio de Naray et al. 
(2010) shown as crosses. All three models have the same $M_{200}=3\times 10^
{12}~M_{\odot}$ but different central densities; $\rho_{0}=5.0,Q_{0}=2.1
\times 10^{-4}$ (red); $\rho_{0}=0.1,Q_{0}=1.4\times 10^{-7}$ (green); $
\rho_{0}=0.01,Q_{0}=1.8\times 10^{-9}$ (blue).}
\end{figure}

\clearpage

\begin{figure}
\plotone{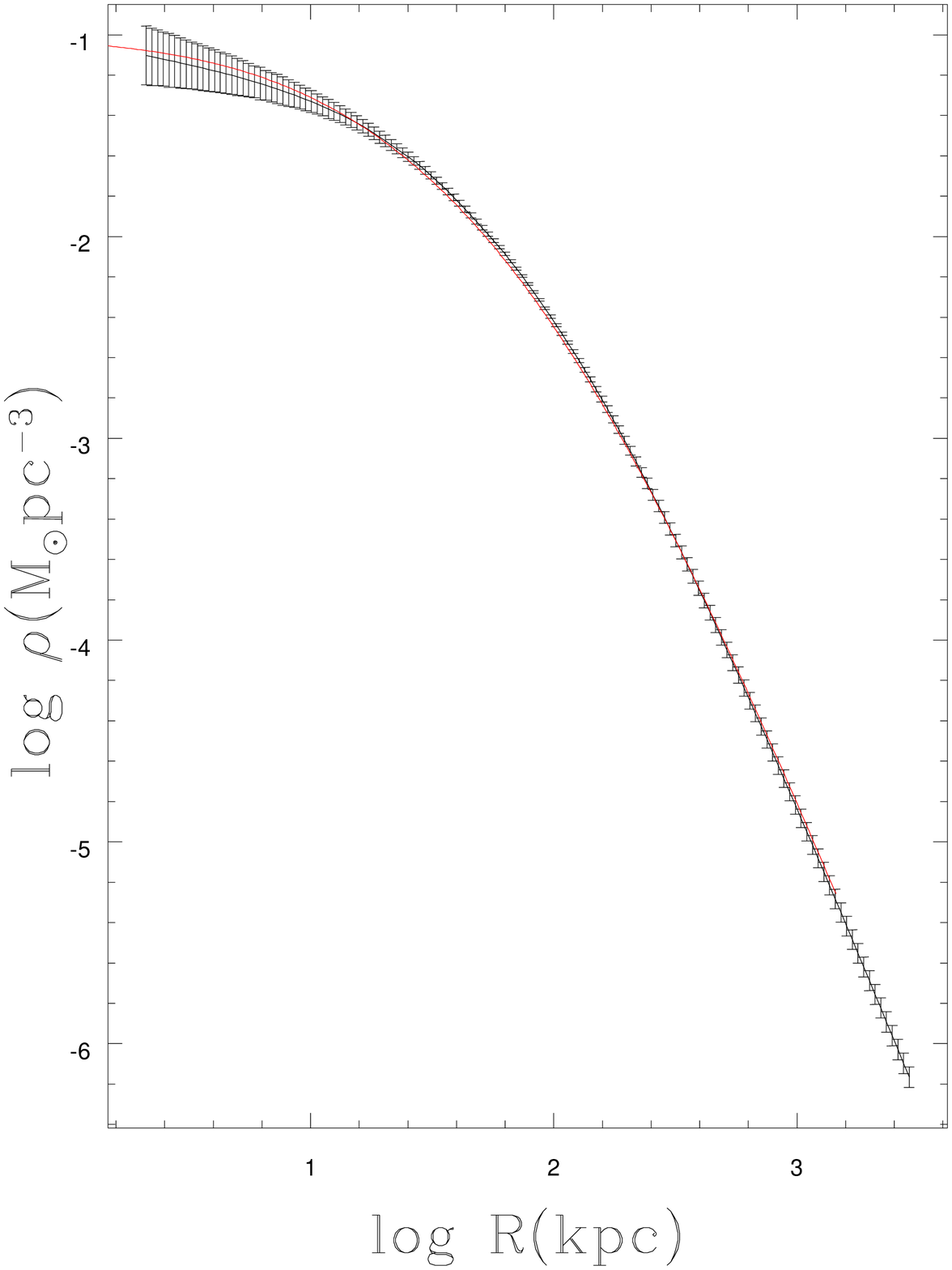}
\caption{A model (red line) with $\rho_{0}r_{0}=2350~M_{\odot}pc^{-2}$ fit to 
the observations of Abell 
611 (Newman et al. 2009). The parameters are $\rho_{0}=0.1,~f=9,~Q_{0}=6.7
\times 10^{-11}~\&~log M_{200}=14.57~M_{\odot}$  } 
\end{figure}


\begin{references}

\reference{}
Amorisco, N. C., \& Evans, N. W. 2011, \mnras, 411, 2118

\reference{}
Binney, J., \& Tremaine, S. 1987, Galactic Dynamics, (Princeton: Princeton 
University Press)


\reference{}
Burkert, A. 1995, \apj, 447, L25

\reference{}
Burkert, A. 2000, \apj, 534, L143

\reference{}
Dalcanton, J.J., \& Hogan, C.J. 2001, \apj, 561, 35

\reference{}
Donato, F., Gentile, G., Salucci, P., Frigerio Martins, C., Wilkinson, M. I., 
Gilmore, G., Grebel, E. K., Koch, A., \& Wyse, R. 2009, \mnras, 397, 1169

\reference{}
Hartwick, F. D. A. 2008, \apj, 674, 220

\reference{}
Hogan, C. J. \& Dalcanton, J. J. 2000, Physical Review D, 62, 063511

\reference{}
Kormendy, J., \& Freeman, K. C. 2004, IAU Symposium 220, 377

\reference{}
Kuzio de Naray, Rachel, Martinez, Gregory D., Bullock, James S., Kaplinghat, 
Manoj 2010, \apj, 710, L161

\reference{}
Loeb, Abraham, \& Weiner, Neal 2011, Physical Review Letters, 106, 171302

\reference{}
McGaugh, S. S., de Blok, W. J. G., Schombert, J. M., Kuzio de Naray, R., Kim, 
J. H. 2007, \apj, 659,149

\reference{}
Navarro, Julio F., Frenk, Carlos S., \& White, Simon D. M. 1997, \apj, 490, 493

\reference{}
Navarro, Julio F., Ludlow, Aaron, Springel, Volker, Wang, Jie, Vogelsberger, 
Mark, White, Simon D. M., Jenkins, Adrian, Frenk, Carlos S., \& Helmi, Amina 
2010, \mnras, 402, 21

\reference{}
Newman, Andrew B., Treu, Tommaso, Ellis, Richard S., Sand, David J., Richard, 
Johan, Marshall, Philip J., Capak, Peter, \& Miyazaki, Satoshi 2009, \apj, 
706, 1078

\reference{}
Newman,Andrew B., Treu, Tommaso, Ellis, Richard S., \& Sand, David J. 2012, 
\apj, in press, astro-ph:1209.1392v1  

\reference{}
Salucci, Paolo, Wilkinson, Mark I., Walker, Matthew G., Gilmore, Gerard F., 
Grebel, Eva K., Koch, Andreas, Frigerio Martins, Christiane, \& Wyse, Rosemary
F. G. 2012, \mnras, 420, 2034

\reference{}
Spergel, D. N. \& Steinhardt, P. J. 2000, Physical Review Letters, 84, 3760

\reference{}
Steinmetz, M. \& Navarro, J. F. 1999, \apj, 513, 555

\reference{}
Strigari, Louis E., Bullock, James S., Kaplinghat, Manoj, Diemand, Juerg, 
Kuhlen, Michael, \& Madau, Piero 2007, \apj, 669, 676

\reference{}
Strigari, Louis E., Bullock, James S., Kaplinghat, Manoj, Simon, Joshua D., 
Geha, Marla, Willman, Beth, \& Walker, Matthew G. 2008, Nature, 454, 1096

\reference{}
Taylor, James E., \& Navarro, Julio F. 2001, \apj, 563, 483

\reference{}
Walker, Matthew G., McGaugh, Stacy S., Mateo, Mario, Olszewski, Edward W., 
Kuzio de Naray, Rachel 2010, \apj, 717, l87

\reference{}
Walker, Matthew G., \& Peñarrubia, J. 2011, \apj, 742, 20

\reference{}
Walker, Matthew G., Mateo, Mario, Olszewski, Edward W., Peñarrubia, Jorge, Wyn
Evans, N., \& Gilmore, G. 2010, \apj, 710, 886

\end{references}
\end{document}